\documentclass[journal=ancac3,manuscript=article]{achemso}

\usepackage{chemformula} 
\usepackage[T1]{fontenc} 

\usepackage{hyperref}
\usepackage{units}
\usepackage{xcolor}
\usepackage{cleveref}

\newcommand*{\citen}[1]{%
  \begingroup
    \romannumeral-`\x 
    \setcitestyle{numbers}%
    \cite{#1}%
  \endgroup   
}

\author{D. Sch\"onke}
\affiliation[Johannes Gutenberg-Universit\"at Mainz]{Institute of Physics, Johannes Gutenberg-Universit\"at Mainz, 55099 Mainz, Germany}
\author{R. M. Reeve}
\affiliation[Johannes Gutenberg-Universit\"at Mainz]{Institute of Physics, Johannes Gutenberg-Universit\"at Mainz, 55099 Mainz, Germany}
\alsoaffiliation{Graduate School of Excellence Materials Science in Mainz (MAINZ),
	Staudinger Weg 9, 55128 Mainz, Germany}
\email{reeve@uni-mainz.de}
\author{H. Stoll}
\affiliation[Johannes Gutenberg-Universit\"at Mainz]{ 
Institute of Physics, Johannes Gutenberg-Universit\"at Mainz, 55099 Mainz, Germany
}
\alsoaffiliation{Max Planck Institute for Intelligent Systems, Heisenbergstrasse 3, 70569 Stuttgart, Germany}
\author{M. Kl\"aui}
\affiliation[Johannes Gutenberg-Universit\"at Mainz]{ 
Institute of Physics, Johannes Gutenberg-Universit\"at Mainz, 55099 Mainz, Germany
}
\alsoaffiliation{Graduate School of Excellence Materials Science in Mainz (MAINZ),
Staudinger Weg 9, 55128 Mainz, Germany}

\title[Competing magnetic states and switching pathways in curved nanowires]
  {Quantification of competing magnetic states and switching pathways in curved nanowires by direct dynamic imaging}


\begin{document}

%
%
%

\begin{abstract}
For viable applications, spintronic devices based e.g. on domain wall motion need to be highly reliable with stable magnetization states and highly reproducible switching pathways transforming one state to another. The existence of multiple stable states and switching pathways in a system is a definitive barrier for device operation, yet rare and stochastic events are difficult to detect and understand. We demonstrate an approach to quantify competing magnetic states and stochastic switching pathways based on time-resolved scanning electron microscopy with polarization analysis, applied to the technologically relevant control of vortex domain wall chirality via field and curvature in curved wires.
While being a pump-probe technique, our analysis scheme nonetheless allows for the disentanglement of different occurring dynamic pathways and we can even identify the rare events leading to changes from one magnetization switching pathway to another pathway via temperature- and geometry-dependent measurements. The experimental imaging is supported by micromagnetic simulations to reveal the mechanisms responsible for the change of the pathway. Together the results allow us to explain the origin and details of the domain wall chirality control and to quantify the frequency and the associated energy barriers of thermally activated changes of the states and switching pathways. 
\end{abstract}

\section{Introduction}
For technological spintronic applications such as magnetic memories~\cite{parkin2008magnetic}, a high thermal stability of the magnetic states and control of their dynamics is required. In order to store or process data in such a device it is necessary to switch the magnetization configuration of the device with a high reliability between different states, which are themselves stable and thus resilient against thermal perturbations. In the simplest case data is stored in a single domain bit whose orientation can be switched by e.g. fields or currents. Here switching can occur by coherent~\cite{schumacher2003phase,schumacher2008biased} or incoherent~\cite{ledermann1995experimental,oliveira2008magnetization} magnetization rotation. However, a number of proposed applications are based on more complex magnetic states involving e.g. discs with vortices or rings and wires with domain walls and for such mesoscopic structures switching in general occurs via more complex processes like the nucleation of domains and domain wall motion~\cite{klaeui2008head,reeve2019magnetization}. For such devices, in particular the stray field interaction between adjacent magnetic structures can also have an impact on the stability of the domains and the switching process reliability~\cite{obrien2009near,sampaio2013coupling,laufenberg2006quantitative,hayward2010direct, hayward2010pinning}. In general for devices based on domain wall (DW) propagation, various approaches to realize DW motion exist including manipulation via current-induced torques on the magnetization~\cite{berger1978low,boulle2011current}. However the most widely studied approach that is also employed in commercially available sensors\cite{diegel2009anew} remains nucleating and moving DWs via applied magnetic fields~\cite{ono1999propagation,hayashi2007direct}. There is also the possibility of so-called DW automotion without a driving field or current in wires with a width or curvature gradient~\cite{mawass2017switching,yershov2018geometry}. Here the DWs travel along the energy gradient to minimize their total energy, which can provide a relevant contribution to the dynamics and switching. An understanding of magnetization switching pathways, DW motion and spin-structure interactions is crucial for various applications such as proposed spintronic memory devices~\cite{parkin2008magnetic}, DW logic~\cite{allwood2005magnetic,omari2019toward} and sensors~\cite{diegel2009anew,borie2017reliable}. In general a high reliability of the magnetization switching pathways and a high stability of the related spin structures is a major criterion for the functionalization of these systems, however so far this could only rarely be assessed in the experimental investigations. One example of a relevant potential application is the case of chirality-encoded logic where the chirality of the propagating vortex domain walls must be controlled with a high-degree of reproducibility~\cite{omari2019toward}. Several approaches to preferentially select a given chirality have been proposed, e.g. using electric fields~\cite{chen2016electric,beardsley2017deterministic} and tailoring of the geometry and applied field~\cite{li2018dynamics} using edges~\cite{omari2015ballistic}, notches~\cite{brandao2014control} or using wire ends as domain wall rectifiers~\cite{wilhelm2009vortex}, but the reliability of the approaches and thus their suitability has mostly not been quantified.
Apart from the technological potential there is also a high scientific interest in understanding magnetic states and their dynamics. In particular the investigation of changes of magnetic states and switching pathways due to thermal activation allows for the quantification of energy barriers, which are so far unknown but also crucial to assess the reliability. Especially curved nanowires are a strongly investigated and technologically relevant system showing complex DW dynamics like oscillating DW velocity~\cite{bisig2013correlation} even below the Walker breakdown~\cite{schryer1974the} and new chiral effects induced by geometry~\cite{volkov2019experimental}.

In order to characterize magnetization switching and DW motion, magnetoresistance measurements~\cite{hayashi2007direct} are frequently employed, but this approach provides limited information about the exact spin structure and switching process. This detailed spin configuration can be obtained by high-resolution magnetic imaging. By using pump-probe techniques, even dynamic imaging with high temporal resolution is possible~\cite{raabe2005quantitative}. However in general, pump-probe imaging techniques average over many cycles, so that only reproducible magnetization dynamics will give a significant signal. As a result, rare events with a low probability are not seen and different switching pathways with comparable probability are superimposed in the final images. Therefore the data are difficult to interpret and are prone to wrong interpretations. Nevertheless, such rare and competing switching pathways are exactly the essential processes in any application since they limit device reliability and can ultimately prevent a device from coming to the market. Here it is of paramount importance to develop techniques that can quantify and understand exactly these processes, while having high time resolution so that one can identify these competing magnetic states and switching pathways and reveal the origin of changes between them.

In this work we employ a specially adapted time-resolved scanning electron microscope with polarization analysis~\cite{schoenke2018development,froemter2016time} (SEMPA) to investigate the magnetic states and the magnetization dynamics of magnetic half-ring (HR) pairs as a prototypical device relevant geometry. We show that through careful analysis of the data we are able to extract details of competing dynamic switching processes from this pump-probe imaging technique. We identify different occurring dynamic pathways, quantify their probabilities and control their prevalence via temperature. Finally this allows us to quantify the energy barrier to change between the different pathways. We perform micromagnetic simulations to understand the magnetization switching within different pathways and in particular reveal the key processes leading to the changes of the pathway that are thus governing the operation reliability.

\section{Results and discussion}

\begin{figure}[H]
\centering
	\includegraphics[width=0.85\linewidth]{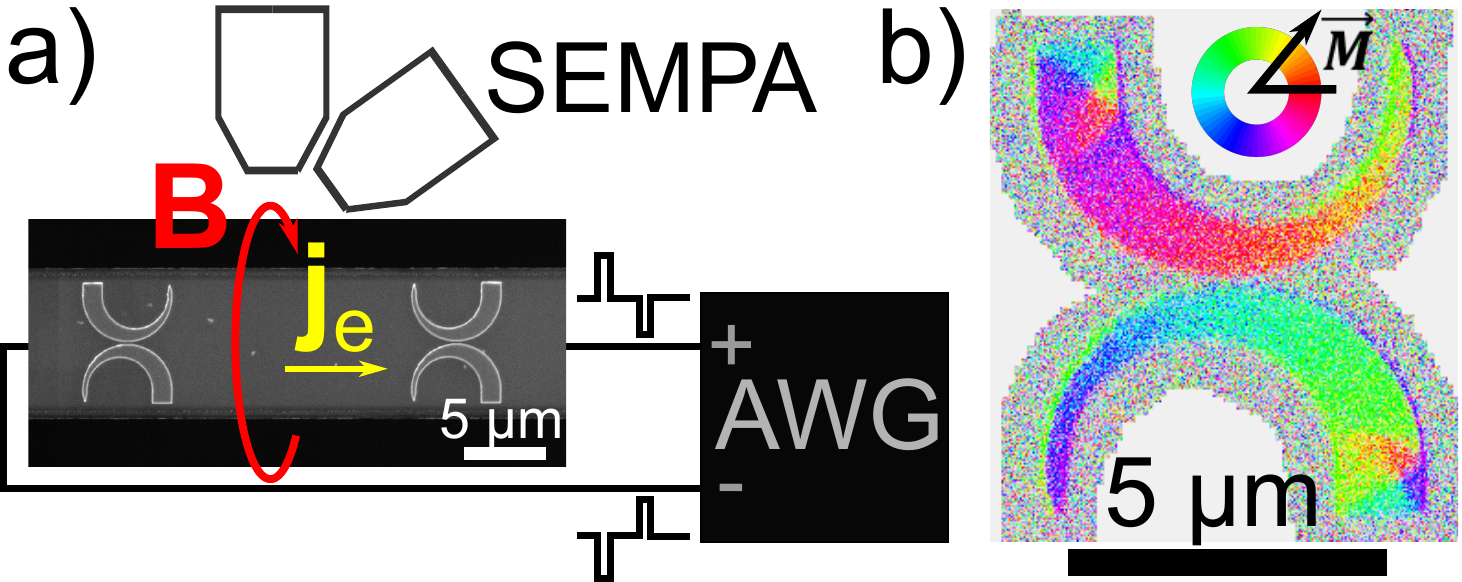}{\centering}
	\caption{\textbf{a)} SEM image of two Py HR pairs with different symmetry on a stripline. The experimental SEMPA setup and the generation of the magnetic field pulses $\vec{B}$ with current pulses $\vec{j_e}$ using an arbitrary waveform generator (AWG) are indicated.  \textbf{b)} Color-coded static SEMPA image of HR1 before excitation. A vortex end domain structure is formed in the wider end of each HR.}
	\label{Experiment}
\end{figure}

Our magnetic structure is composed of two mesoscopic Ni$_{80}$Fe$_{20}$ HRs with a width gradient as can be seen in \autoref{Experiment}a). The fabrication and sample structure of the four half rings studied in detail (HR1--HR4) is described in the methods \autoref{Methods}.

\autoref{Experiment}b) shows a static color-coded magnetic image of HR1 before applying a field to it. The wider end of the wires is rectangular and forms a flux closure vortex end domain structure to minimize the stray field~\cite{gomez1999domain}, whereas the other end of the wire is tapered and does not support such a spin structure~\cite{hayashi2006influence}. The HRs exhibit two principal states at remanence. The first one, which is shown in \autoref{Experiment}b), is a quasi-uniform state where, aside from the end vortex, the magnetization tends to follow the curvature of the ring from one side to the other as a result of shape anisotropy. The second is a half-onion state~\cite{rothman2001observation}, where a DW is additionally present in the structure. The states can be switched between each other via the nucleation (or annihilation) of DWs, which due to the curvature can be generated in the center of the HR via a transverse in-plane field pulse~\cite{klaeui2008head}. Based on the chosen width and thickness, vortex domain walls (VWs) are the stable DW configuration~\cite{laufenberg2006quantitative}. We measured HR pairs with two different width gradients and two different symmetric arrangements: The HRs in each pair are oriented with their vertices towards each other, so that the width gradient is either along the same direction as seen in the pair on the left of \autoref{Experiment}a) (point symmetry: HR1 and HR3) or the opposite direction as seen in the right of \autoref{Experiment}a) (axis symmetry: HR2 and HR4). HR1 and HR2 have the same width gradient, as do HR3 and HR4 (details see methods \autoref{Methods}). The structures are placed on top of a stripline, which generates a field along the axis connecting the vertices of both HRs when a current is flowing (\autoref{Experiment}a)). We continuously excite the system with two subsequent short current pulses with alternating direction as shown schematically on the figure. 
The delay between the pulses is \unit[75]{ns}, the period length \unit[150]{ns}. During the excitation we image the magnetization dynamics of the switching and DW nucleation processes using time-resolved SEMPA as described in Refs.~\citen{froemter2016time,schoenke2018development}.

\subsection{Curvature dependent chirality of nucleated vortex walls}

\begin{figure}[H]
\centering
	\includegraphics[width=0.69 \linewidth]{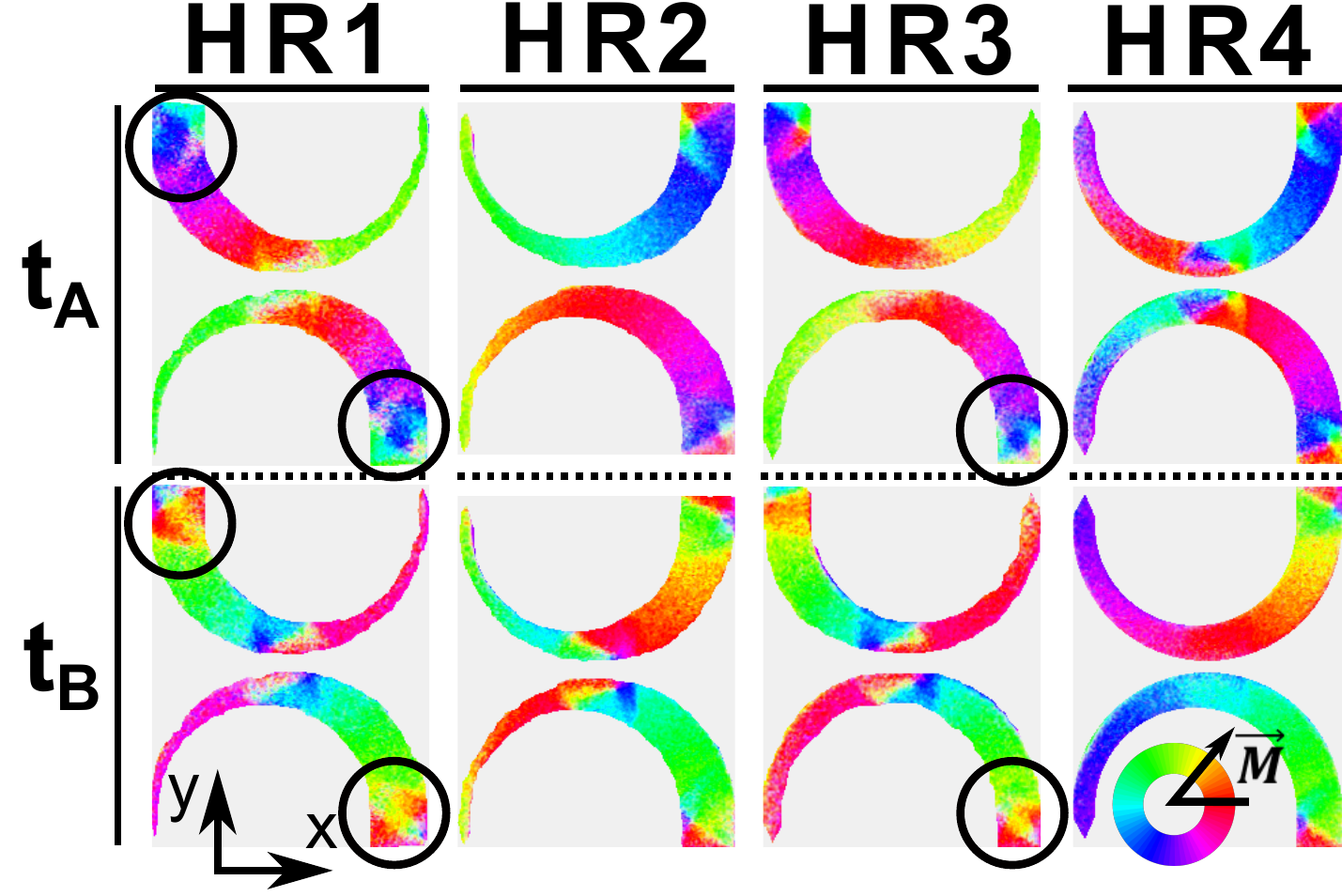}{\centering}
	\caption{Magnetic steady--state configuration for all HRs after the $-B_{y}$ pulse ($t_A$) and after the $+B_{y}$ pulse ($t_B$) obtained during the dynamic imaging with the application of the pulse sequence. The $x$/$y$ axes are shown in the bottom left corner. Some of the end domains show unphysical magnetization configurations which can be interpreted as a superposition of multiple states (encircled). The  nucleated center VWs after the opposite magnetic field pulses at times $t_A$ and $t_B$ display clearly different contrast indicating different probabilities of their nucleation process.}
	\label{steadystate}
\end{figure}

First we demonstrate the technologically relevant high reliability of VW chirality control in our geometry by looking at the remnant state of the magnetization during the periods of the pump-probe excitation cycle when the field is not applied. The chirality of the nucleated VW strongly depends on the combination of the HR curvature and the direction of the nucleating B-field. This can be seen in \autoref{steadystate}, which presents the magnetic steady--state configuration that is measured in the short time frame between the pulses with opposite field direction for the different samples during the dynamic imaging process. The experimental magnetization images of the samples with axis symmetry (HR2 and HR4) show the same chiralities for the nucleated VWs at the vertices of both the upper and lower HRs. Here VW nucleation occurs only after one of the field pulses. In the case of point-symmetrically arranged HRs (HR1 and HR3) the chiralities of the generated VWs are opposite in the two HRs in each pair. For the lower HR which has a fixed orientation the chirality is for all structures the same for a given field direction. Comparing the center VWs of HR1 or HR3 in the time frames after the $-B_y$ pulse ($t_A$) and after the $+B_y$ pulse ($t_B$) in \autoref{steadystate}, it can be seen that the VW chirality changes with the direction of the field pulse. In the next section we will see this confirmed also for HR4. The experimental results could be supported by micromagnetic simulations (for details see methods \autoref{Methods}), showing the same correlation of the nucleated VW chirality, geometry and magnetic field direction. 

From the time-resolved imaging and simulations we also determine the dynamics of the magnetic switching. 
A systematic summary of the VW chiralities and types (head-to-head or tail-to-tail) depending on the HR curvature and the field direction for all measured and simulated HR pair geometries, as well as the full details switching process, is given in the supplemental information (section S1-S3, videos in S6).
The process of the VW formation occurs via nucleation and annihilation of multiple vortices, however only the vortex with the preferred chirality due to the curvature remains and is stable after the pulse. In the dynamic imaging we observe pinning of the wall after the pulse is switched off and subsequent expulsion of the wall at the inner edge of the HR during the application of the next, reverse, pulse.

This leads to the state following successive pulses alternating between one with a DW in the center and one with no DW following annihilation, as seen in \autoref{steadystate} for HR2 \& HR3. However, while this dynamic scheme is generally followed, our results reveal that it is not reproduced with 100\% reliability. Firstly, in some simulations (see e.g. HRS1 and HRS2 videos in supplemental material section S6), instead of DW pinning we observe DW motion towards the narrower end of the HR driven by the width gradient. In this case, the following pulse can nucleate a new VW. Secondly, taking a closer look at the dynamically obtained experimental images in \autoref{steadystate} we can identify some puzzling features. One discrepancy can be noted with some of the end domain structures. If we compare the steady--state image of HR1 in \autoref{steadystate} obtained by pump probe imaging between pulses over many pulse excitation cycles to the corresponding completely static image in \autoref{Experiment} obtained without any excitation, we see strong differences. The encircled states in HR1 and also HR3 in \autoref{steadystate} are not the expected vortex structure and at first sight it looks like an unexpected high energy configuration that we cannot reproduce by micromagnetics. Additionally there is strongly differing contrast of the nucleated VW after the negative magnetic field pulse compared with the contrast after the positive pulse, as can be seen e.g. by comparing the frames $t_A$ and $t_B$ of HR1 \& HR3 in \autoref{steadystate}. We postulate that these unexpected features are a result of the employed dynamic pump-probe imaging scheme that integrates over many cycles of the excitation. If the HRs exhibit competing magnetic states and switching pathways then these are superimposed in the resulting images, resulting in either reduced contrast or unphysical apparent magnetic configurations. In conventional analysis further details concerning the origin and prevalence of such different configurations and dynamics are not extracted from such pump-probe imaging. However in the next section, we show that with our analysis method it is indeed possible to characterize such stochastically changing dynamic pathways which are crucial to the device operation reliability.

\subsection{Switching pathways}

\begin{figure}[H]
\centering
	\includegraphics[width=0.85\linewidth]{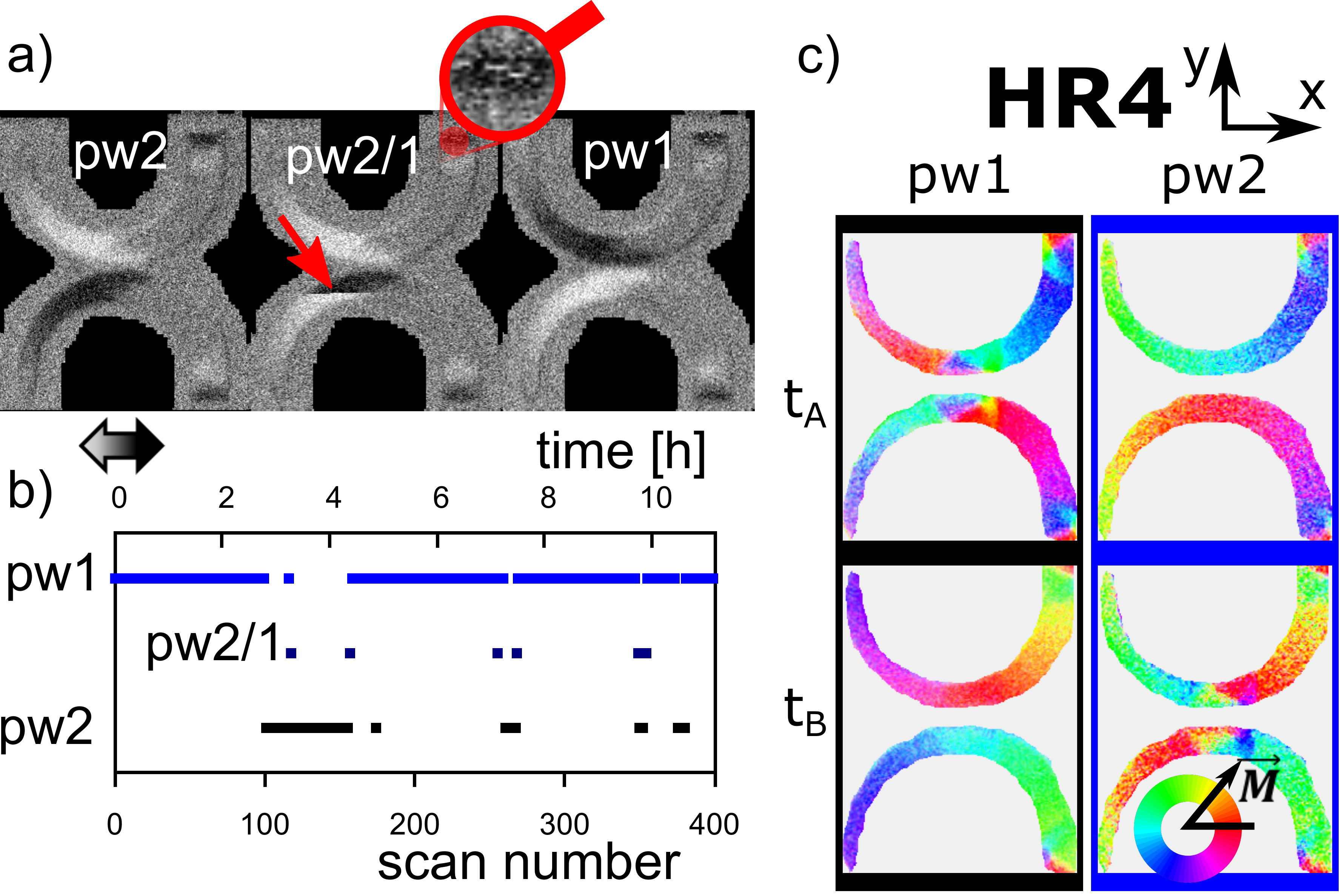}{\centering}
	\caption{\textbf{a)} Single scan images of HR4. The SEMPA images show the integrated magnetization over the full excitation period. The three shown images (pw2, pw2/1 and pw1) are subsequent scans. Comparing the left scan with the right scan, it is obvious that the magnetization of the HRs has changed in the narrower part, indicating a change back from pathway 2 (pw2) to pathway 1 (pw1). During the acquisition of the lower HR in scan pw2/1, a change between the two switching pathways took place from pathway 2 to pathway 1, leading to the abrupt change in contrast, as indicated by an arrow. Abrupt contrast changes in the end vortex can be seen in the magnified image section. \textbf{b)} The graph shows the changes between pathways for each of the 400 single scans, which were acquired over a total time of about \unit[11]{h}. \textbf{c)} The steady--state magnetization SEMPA images of the two possible pathways just before the field pulses after disentangling both pathways (c.f. \autoref{steadystate})}
	\label{HR1pathways}
\end{figure}

To envision devices based on DW dynamics, the magnetic states and the switching pathways need to be controllable with high reliability. Furthermore, if multiple pathways exist, one must have means to analyze them individually as well as the probabilities of changing between the switching pathways. In order to investigate the reliability of VW nucleation dynamics in HR1-4 we take advantage of the fact that the SEMPA acquisition consists of the addition of many subsequent high speed scans of the field of view, which individually provide a wealth of information that is typically not used: While the final videos show all occurring switching pathways superimposed in one movie, the single scans capture details of single pathways that are not changing over the millisecond timescale used as the acquisition time for a pixel in an individual scan. To see this, we next look at images that are the time integration over a whole excitation period for a particular scan across the sample. When we plot this average magnetization integrated over the whole excitation period, we see that there are two switching pathways superimposed. The acquisition time for a single scan is lower than the typical time between changes of the switching pathways for HR4, so that we can disentangle both pathways as shown in \autoref{HR1pathways}. 

Here the SEMPA single scan images (\autoref{HR1pathways}a)) which show the magnetization integrated over a full excitation period for successive scans of the sample show opposite $x$-direction contrast in the narrower part of the wire for different scans, indicating different switching pathways at these different times. During the acquisition of the middle scan of \autoref{HR1pathways}a) a change between the pathways occurred, as visible from the abrupt change in contrast indicated by the red arrow. The probability and time scale of these two switching processes is shown in \autoref{HR1pathways}b). After determining which scans belong to which pathway, the scans can be added together to make two final movies which each show a single pathway. Now we get two separate movies and can analyze the two pathways individually. \autoref{HR1pathways}c) shows, in analogy to \autoref{steadystate}, the dynamically measured steady--states in HR4 after the $-B_y$ pulse ($t_A$) and after the $+B_y$ pulse ($t_B$) for the two pathways (pw1, pw2), revealing new information that cannot be obtained from conventional imaging and analysis: First, for pathway 2 (pw2) we see VW nucleation by the negative field pulse, which is not obvious in \autoref{steadystate} (HR4 $t_B$) due to the lower probability of this pathway. Second, as described before, we see that a vortex wall only remains at the HR center after one of the two pulses for a given pathway and that the pulse after which this occurs switches for the two pathways. This allows for a more detailed comparison of both pathways as well as additional information such as the determination of the VW velocities, which is provided in the supplemental material section S2. 

\begin{figure}[H]
\centering
	\includegraphics[width=0.69 \linewidth]{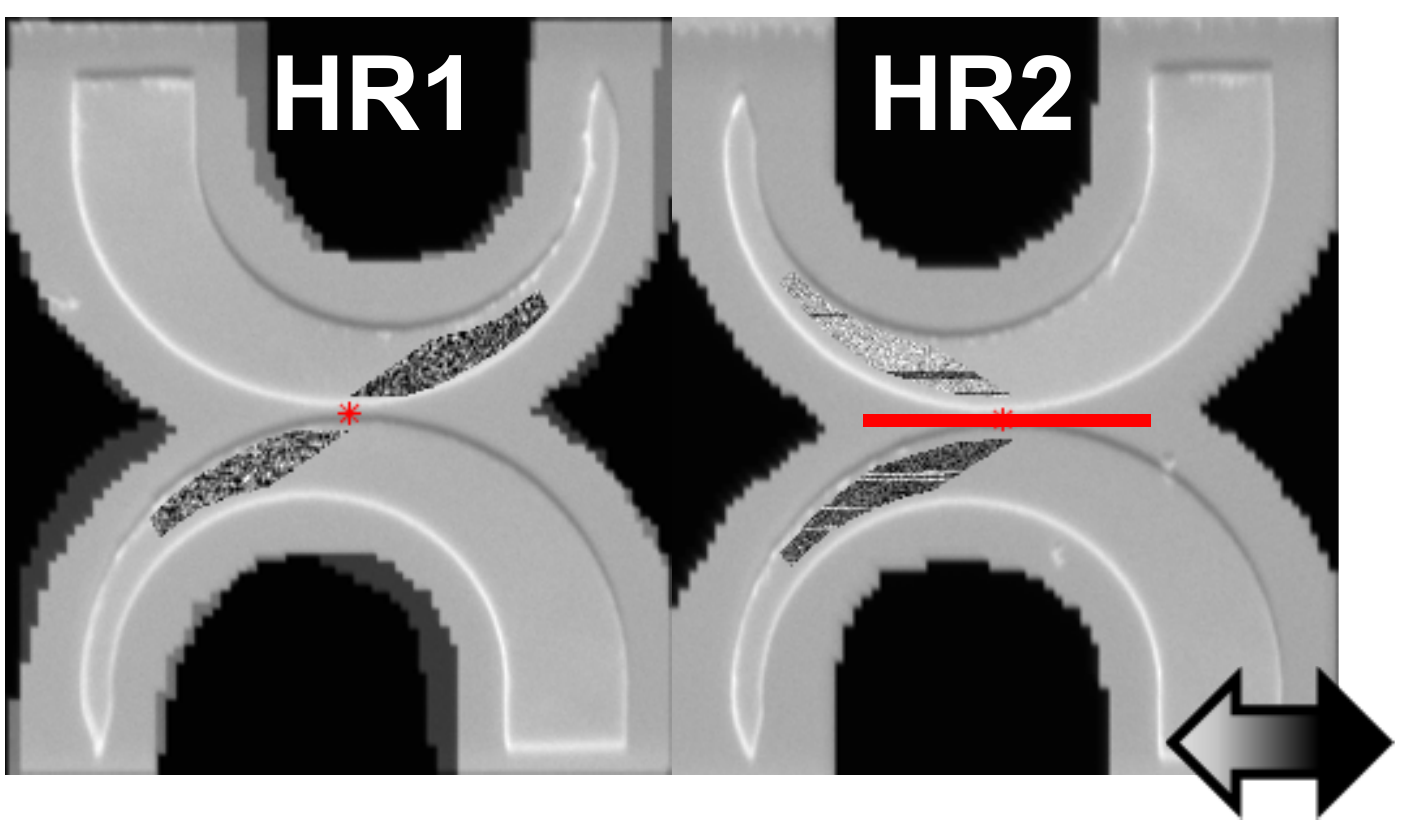}{\centering}
	\caption{Topographic image of HR1 and HR2 with overlay of the SEMPA single scan $x$ magnetization images of HR1/2 integrated over the whole excitation period. For HR2, where the HRs are arranged in axis symmetry, the changes between the pathways are much less frequent than for the point symmetric HR1, where the time between the change in the pathway is of the same order as the pixel dwell time of the scanning electron beam (\unit[2]{ms}). The symmetry point/ axis are indicated in red.}
	\label{HR56_orientation}
\end{figure}

Close inspection of the end vortices also reveals statistically significant rows of white pixels in the otherwise dark end domain (magnified in \autoref{HR1pathways}a)) which can be attributed to switching of the end vortex chirality for different excitation cycles as well. However, the timescale and probabilities of the possible end domain states differ from the two switching pathways and are therefore not correlated to the changing between the two pathways of the VW nucleation that we discussed above. Also note that the individual determination of switching pathways is limited by the necessary integration time. For instance at room temperature for HR1 and HR2 the allocation of the individual scans to a particular switching pathway is not possible because of a much higher pathway changing frequency. Here the changing between pathways is still present but manifests itself as a streaky magnetization pattern in the single scans, as can be seen in the images in  \autoref{HR56_orientation}. However we can still identify that the changing of the pathways in the upper and lower HR is coupled, with no cases observed when only one DW was present in the HR center. Having established the presence of multiple pathways, the final step is to identify the origin of the changes between the switching pathways: To this end in the next section the dependency of pathway switching on the orientation of the adjacent HRs and on temperature is studied to identify the mechanism that leads to the change in switching pathway.

\subsection{Prevalence of different pathways}
Comparing the single scan SEMPA $x$-component images of HR1-4 in \Cref{HR1pathways,HR56_orientation}, it is obvious that the changes from one pathway to the other occur more often for the HR pairs which are arranged in a point-symmetric manner (HR1, HR3). While for HR4 the time between the change of pathways is on the order of minutes or even hours, as shown in \autoref{HR1pathways} (the acquisition time for a single scan image is \unit[100]{s} followed by a short processing time required by the software), in HR3 at the same temperature (\unit[340]{K}) the time between the events that change the pathway is only a few ms (c.f. supplemental materials Fig. S3). 
The same holds for HR1 and HR2 with a higher width gradient. Here the overall frequency of pathway change is strongly increased (\autoref{HR56_orientation}), so that in HR1 the streaky pattern is no longer visible and the changing between the pathways now only resembles enhanced noise. In general, the rare event leading to the change in pathway occurs more often in HR1 and HR2, where the width gradient is higher than for HR3 and HR4. In the micromagnetic simulations HRS1\&2 (provided in the supplemental materials section S6) VW nucleation after both field pulses in the same period is seen as a result of automotion of the VW towards the narrow end of the HR as the field pulse is removed. This provides hints for the mechanism for pathway changing in the experiment: While usually the VW is pinned in the center in the experimental system, thermal activation during the relaxation can on rare occasions overcome the dynamic pinning and yield rare automotion events that then lead to the change in the pathway. A quantitative analysis of even this rarely occurring mechanism is possible by counting the number of pathway changes in a certain time, as enabled by our analysis method. This is supported by measurements of the same HR at different temperatures, revealing more frequent changes of the pathway at higher temperatures (details, see supplementary materials section S4) and showing that different pathways can be individually detected by lowering the temperature so the separation between different pathways becomes possible.

While the results reveal stochastic changes to the switching pathway they nonetheless show reliable control of the chirality of the nucleated VW. Depending on the curvature and the relative direction of the applied field, VWs with a specified chirality can be nucleated independent of the end vortex chirality. The underlying mechanism is not simple motion of the end vortex, but rather a complex successive nucleation of vortices and antivortices is revealed. Thus the end vortex plays a role in determining the switching dynamics between a state without a VW in the HR center and the state with a VW, but it does not change the resulting spin structure chirality. The end vortices are stabilized by stray field minimization. Nevertheless, the end vortex chirality is found to change, as revealed by white stripes in the dark end domains in \autoref{HR1pathways}a). This shows the limitations of a previously suggested approach for reliable spin structure rectification by end domains~\cite{wilhelm2009vortex}. Also this explains the appearance of the apparent complex end domain patterns for HR1 and HR3 seen in \autoref{steadystate} which are not physical and instead correspond to a a superposition of different vortex states with opposite chirality.  
We additionally find that the HRs exhibit multiple switching pathways due to thermally activated changes of the dynamics. 
The frequency of changing between pathways is found to be larger in HRs with a higher width gradient and for the case of point-symmetric orientation. This is consistent with an explanation based on DW automotion away from the center of the wire, as observed in the simulations and described above. A higher width gradient favors automotion of the VW towards the narrower end of the wire to reduce the domain wall energy~\cite{mawass2017switching}. In the videos showing the magnetization dynamics, which average over all excitation periods during the measurement time, automotion is not directly visible. This is to be expected, since the automotion is related to the change in the switching pathway and hence is a relatively rare event and it is not captured in the dynamic movies which require highly reproducible dynamics. Nevertheless we can detect and even count the number of the rare events occurring in our experiment by a thorough examination of the raw data. Thermal activation can release the VW from a shallow pinning potential and enable automotion to set in for a small subset of excitation cycles. We can now determine the energy barrier from the frequency of changes between pathways in temperature-dependent measurements of HR3 yielding a value of about $3\times 10^{-21}$\,J (details, see supplemental materials section S4).
If the DW has overcome this barrier during the dynamics and moved far enough before the next Oersted field pulse starts, it will not annihilate with the newly forming vortex of opposite chirality in the wider part of the wire. The interaction between the DWs that depends not only on the magnetic charge but also on their spin structure can help or impede to overcome this energy barrier (c.f. supplemental materials section S5). 
Therefore as a result of the automotion, the switching pathway has changed. If the automotion only leads to a small displacement of the VW, for instance due to pinning or interaction with the DW in the adjacent wire, a more complicated behaviour, which is a mixture of the other cases, can occur. 

\section{Conclusion}
In summary we have studied competing magnetic states and stochastic switching pathways in a prototype magnetic device consisting of two Permalloy HRs using time-resolved SEMPA imaging and micromagnetic simulations and we have identified the key factors that govern the switching mechanism.
In the particular samples investigated, multiple switching pathways are possible: Thermally activated changes between pathways occur and we identify automotion as the mechanism that mediates transitions from one switching pathway to another. The chosen experimental technique provides particular insights into these thermally assisted switching processes and their reliability. By looking at the individual subsequent scans extensive additional information was gained, including information about less probable pathways and even rare events that lead to the pathway changing, which is hidden in conventional pump-probe acquisition and analysis. This allowed us to analyze the statistics of the mechanisms and find, depending on the orientation of the two HRs, a greater or smaller stability of the switching pathway due to DW interaction that helps or impedes DW automotion along the wire width gradient. Temperature dependent measurements confirm the impact of thermal energy on DW automotion and pathway changes and allow us to quantify the energy barrier between different switching pathways.
The work presents a special approach to determine the reliability of magnetization switching pathways and detect even quantitatively rare events resulting in a change of the dynamics. Further we find that the curvature of the structures, in combination with the applied field direction, determines the chirality of the nucleated VW and we demonstrate the high-degree of reproducibility of preferential chirality DW nucleation via curvature. The combination of an easily manufacturable device design with a highly reliable mechanism for nucleating VWs with the appropriate chirality could be a major step towards production maturity of chirality-based logic devices.

\section{Methods}
\label{Methods}
\subsection{Sample Fabrication}
\label{SampleFabrication}
Ni$_{80}$Fe$_{20}$ (Permalloy/ Py) HRs were produced on a stripline for generating a current-induced Oersted field. The samples were fabricated in a two-step electron beam lithography lift-off process: First the stripline was patterned using a MMA-PMMA double layer resist. The 8\,\textmu m wide stripline consists of a Ta(4\,nm) [CuN(16\,nm)Ta(4\,nm)]x3 stack deposited in a Singulus Rotaris sputter deposition system. This stack was chosen because of its optimized low mean surface roughness of about 0.4\,nm (determined by atomic force microscopy) and appropriate resistance for current pulse injection (approx. 82 Ohm). In a second lithography step the magnetic structures were patterned using a PMMA single layer. 30 nm Py was deposited via electron-beam evaporation in a chamber with a low base pressure of 10$^{-9}$\,mbar. The HR has an outer diameter of 5.5\,\textmu m. The width of the HR is gradually decreased from one rectangular end to the tapered end. The inner radius is 2\,\textmu m. Samples with two different gradients have been investigated. For sample HR1 and HR2 the width in the narrowest part is 300 nm, while for HR3 and HR4 it is 500 nm. The HRs are arranged in axis or point symmetry pairs on the stripline (Fig. 1).

\subsection{Experimental Details}
\label{ExperimentalDetails}

We imaged the magnetization dynamics in the halfrings with a newly developed scanning electron microscope with polarization analysis (SEMPA) system that has a time resolution of less than \unit[2]{ns}~\cite{schoenke2018development}. Two current pulses with opposite polarity through the stripline act on the magnetization by generating a magnetic field of \unit[$\pm$15]{mT}. The field was calculated by detecting the voltages before and after the sample with pick-off tees. The pulse length was \unit[21]{ns} and the rise/fall time \unit[8]{ns} with a period length of \unit[150]{ns}. For SEMPA imaging, the sample has to be kept at a fixed potential, which can be done using a push pull circuit with antisymmetric driving voltage~\cite{froemter2016time}. The acquisition was performed by sampling the period time in 75 equally spaced \unit[2]{ns}-bins, i.e. each detector signal is assigned to one of these time windows. To avoid distortions in the image due to thermal drift we used a high scan speed of 500 pixel per second and scanned the field of view multiple times. During post-processing all the counts for every detector channel and pixel are added after cross-correlation alignment of the pictures. The EHT voltage of the primary electron beam was \unit[7.5]{kV}, the beam current \unit[3]{nA}. To have improved magnetic contrast, the samples were milled in--situ with 1\,keV Ar ions and a thin (\unit[5]{\AA}) iron layer was deposited on top~\cite{vanzandt1991iron}. Variable temperature imaging is performed using a liquid helium flow cryostat thermally coupled to the sample reception. The sample temperature is measured by a Pt100 resistor directly on the sample stage.

\subsection{Simulation Details}
\label{SimulationDetails}
The micromagnetic simulations were performed using the MICROMAGNUM code~\cite{micromagnum}. The initial state was set equal to the observed steady--state before the $+B_y$ field pulse by initializing the magnetic chirality of the HR as well as of the end vortex of the HR. The system was subsequently relaxed until the torques in the system were below a threshold. The dimensions of the simulated HRs are the same as those used in the experiment. The material parameters used for the simulation of the magnetization dynamics are the typical values for
Permalloy: exchange stiffness $A = 1.3 \times 10^{-11}\,\mathrm{J/m}$,
saturation magnetization $M_{\mathrm{S}} = 800 \times 10^3\,\mathrm{A/m}$, damping
parameter $\alpha = 0.008$, no net magnetocrystalline anisotropy, and the cell size of
$5\,\mathrm{nm} \times 5\,\mathrm{nm} \times 30\,\mathrm{nm}$~\cite{mawass2017switching}.

\section{Author information}

\subsection{Author Contributions}
D.S., R.M.R., H.S., and M.K. conceived the project; D.S. prepared the samples; D.S. carried out the measurements; micromagnetic simulations and data analysis are performed by D.S. and R.M.R.; D.S., R.M.R. and M.K. prepared the manuscript; R.M.R., H.S., and M.K. supervised the project. All authors discussed the results.

\subsection{Notes}
The authors declare no competing financial interest.
\begin{acknowledgement}

We acknowledge financial support of the SFB/TRR 173 Spin+X: spin in its collective environment funded by the Deutsche Forschungsgemeinschaft (DFG, German Research Foundation) - TRR 173 - 268565370, as well as the Graduate School of Excellence Materials Science in Mainz (No. GSC266).

\end{acknowledgement}

\begin{suppinfo}

The following files are available free of charge.
\begin{itemize}
  \item Supplemental information (PDF)
  \item SEMPA and simulation movies
  \begin{itemize}
    \item HR1 pw2 150 ns (340K)
  	\item HR2 pw2 150 ns (339K)
  	\item HR3 pw2 150 ns (340K)
  	\item HR3 pw2 150 ns (158K)
  	\item HR4 pw1 150 ns (340K)
  	\item HR4 pw2 150 ns (340K)
  	\item HRS1 pw2 109 ns
  	\item HRS2 pw2 96 ns
  	\item HRS3 pw2 113 ns
  	\item HRS4 pw1 114 ns
  	\item HRS4 pw2 90 ns
  \end{itemize}

\end{itemize}

\end{suppinfo}

\bibliography{references}

\end{document}


\section{Detailed description of vortex wall nucleation process}
\begin{figure}[H]
\centering
	\includegraphics[width=0.85\linewidth]{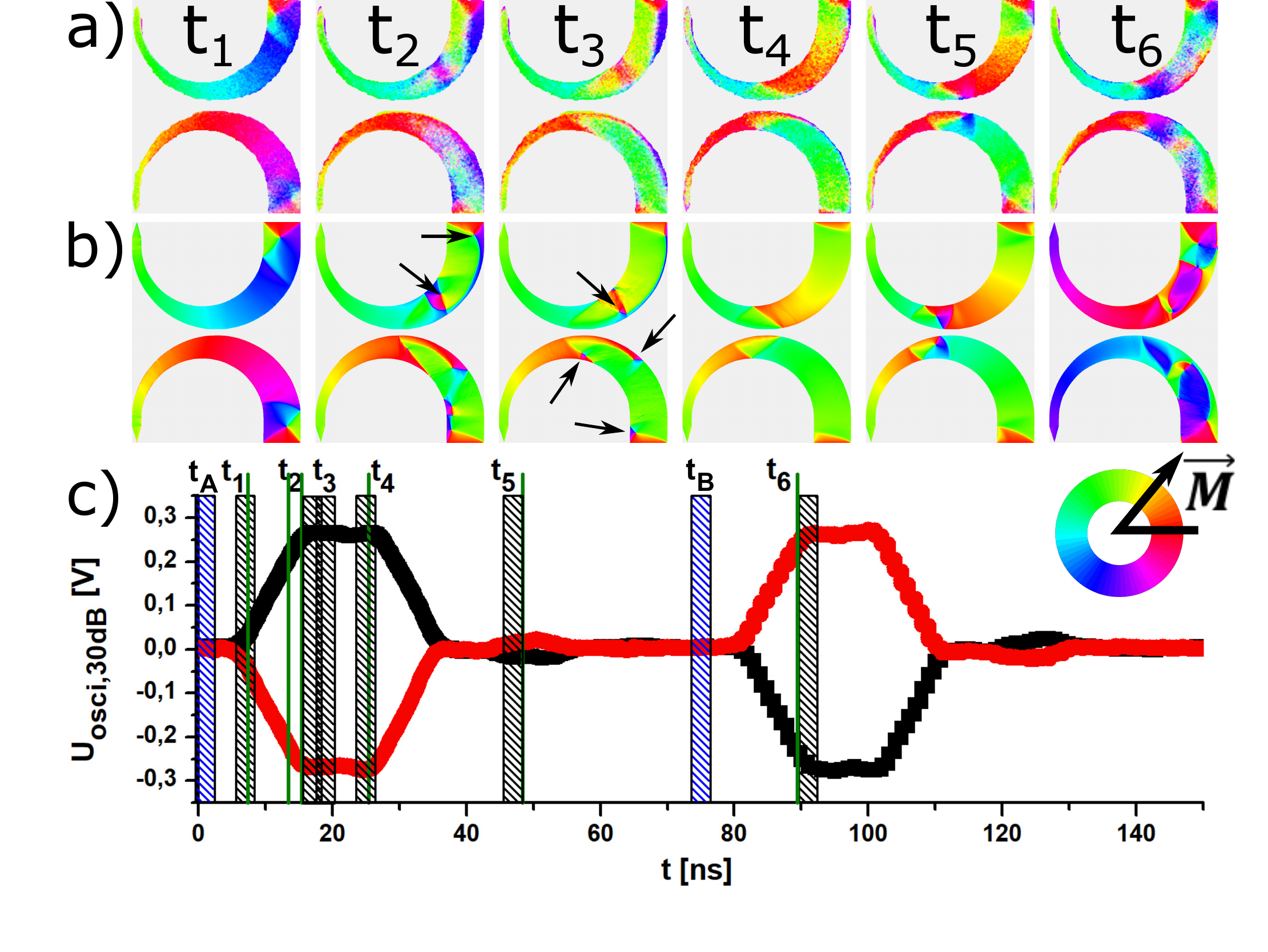}{\centering}
	\caption{Full switching pathway for HR2. 
	\textbf{a)} Experimental SEMPA images at different times $t_1-t_6$.  \textbf{b)} Micromagnetic simulations with B$_{\mathrm{max}}$=\unit[15]{mT}. Black arrows  indicate the three appearing vortices during the magnetization switching in the upper and lower HR. The magnetization direction for both the measurements and the simulations is indicated by the color wheel below.
	\textbf{c)} Pulse shape measured with two \unit[30]{dB} pick-off tees before (black) and after (red) the sample. The black areas indicate the time windows corresponding to a), the green lines the times of the simulation images in b), and the blue areas are the times for the steady--state images in Fig. 2 in the main paper}.
	\label{pathway}
\end{figure}

The process of the nucleation that leads to chirality control is shown in detail for the structure HR2 in \autoref{pathway} where the snapshots of the experimentally imaged magnetization (a)) are compared to micromagnetic simulations (b)) for one geometry. 
We consider the switching during the first pulse with the start magnetization state without a domain wall (DW) in the center of the halfrings (HRs). The description is given for an initial state where the boundary domain of the end vortex points in the same direction as the overall magnetization (\autoref{pathway}: upper HR at $t_1$). The leading edge 90\textdegree\:N\'{e}el-DW of the vortex travels towards the HR center while the domain of the end vortex with the energetically preferred orientation with respect to the field extends. At an angular position of around 45\textdegree\: it splits up into a vortex-antivortex pair~\cite{bisig2015dynamic} ($t_2$ in \autoref{pathway}, where the end vortex and the just nucleated vortex are indicated by black arrows). From the simulations it is revealed that the chirality of the new vortex is opposite to the chirality of the end vortex. 
The leading 90\textdegree\:N\'{e}el-DW again splits into a vortex-antivortex pair with a clockwise chirality, as for the end vortex (also indicated as a black arrow on the upper HR for $t_3$ in \autoref{pathway}b)). This new vortex is then driven along the outer perimeter of the wire towards the vertex of the HR. The end vortex is also squeezed against the outer perimeter of the ring and in some simulations it is observed to unpin from the end and propagate towards the vertex (not shown). In some cases it is then observed that while the third nucleated vortex annihilated again, this initial end vortex can even travel the whole distance to the vertex without being expelled. In each case, however, a single wall remains at the end of the excitation in the center of the wire. The field can expel the vortex core (VC) of this remaining wall at the HR vertex to leave a transverse wall while the pulse is applied ($t_4$ in \autoref{pathway}), but when the VC re-nucleates during the relaxation of the field pulse the vortex has the same chirality ($t_5$ in \autoref{pathway}). 
If the boundary domain of the end vortex points in the opposite direction to the overall magnetization (\autoref{pathway}: lower HR at $t_1$), the vortex generated by the splitting of the leading 90\textdegree -N\'{e}el DW from the end vortex structure is driven along the outer perimeter of the wire towards the center ($t_3$). In some of the simulations in the initial state an additional small vortex of opposite chirality formed at the inner edge of the HR at the wider end at the leading edge of the end vortex. Then this vortex is driven directly to the center of the HR.
Comparing the nucleation mechanism of the upper and the lower HR, it can be seen that the chirality of the prevailing vortex does not depend on the chirality of the initial end vortex, but only on the field direction and the curvature. This is the the reason why for the upper HR the third nucleated vortex prevails while for the lower HR the second one remains and the third one is expelled. Thus one can reliably generate a vortex wall with a defined chirality.

\section{Influence of the end vortex chirality on the VW nucleation}

\begin{figure}[H]
\centering
	\includegraphics[width=0.85\linewidth]{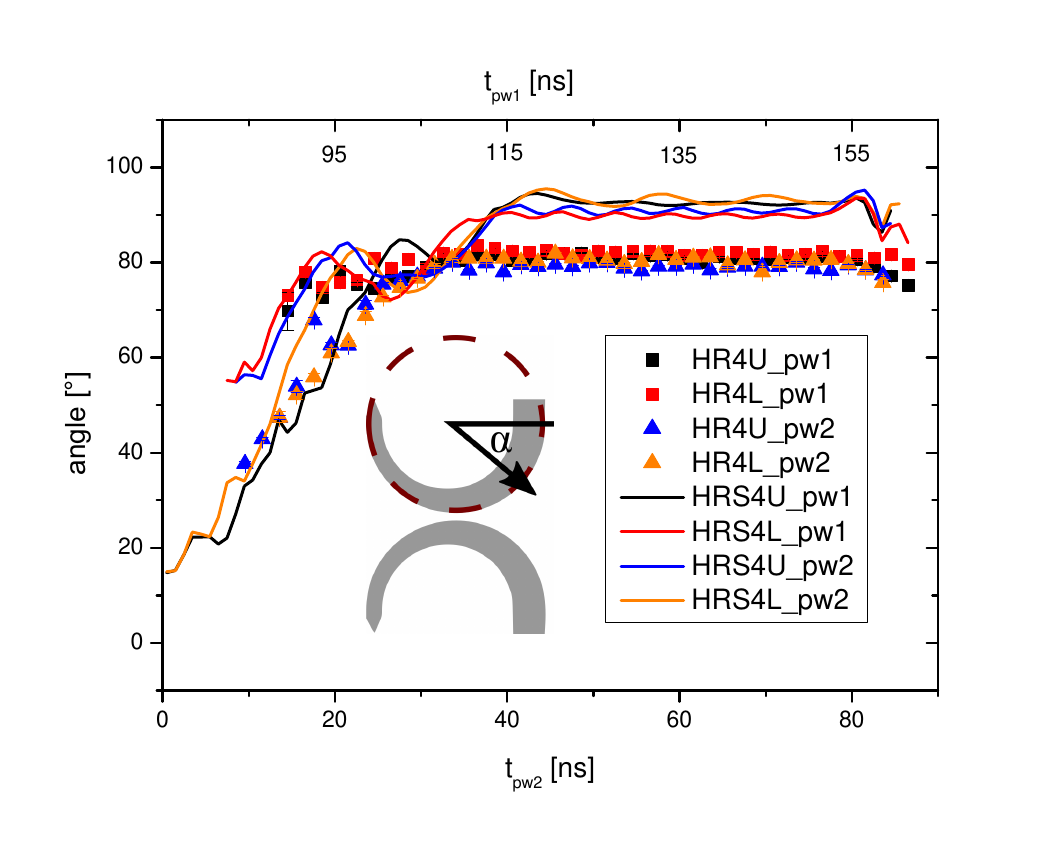}{\centering}
	\caption{Angular position $\alpha$ of the vortex DWs in HR4 comparing both pathways (pw1 and pw2). U is  the upper HR, L the lower one. HR4 refers to the experiment (markers), HRS4 to the corresponding simulation (line plot). The schematic inset shows the polar coordinate system for a HR element with the zero angle always pointing towards the wider end of the wire.}
	\label{HR(S)1pw1pw2a_g}
\end{figure}

The angular position of the remaining VW in the experimental SEMPA images (HR4) and the simulations (HRS4) is shown in \autoref{HR(S)1pw1pw2a_g}. Comparing the movies for the two pathways in HR4, pathway 1 (pw1) and pathway 2 (pw2) for the lower element `L' (HR(S)4L\_pw1 vs. HR(S)4L\_pw2) (\autoref{HR(S)1pw1pw2a_g}), the final vortex reaches the central region of the HR sooner following nucleation in the case of pathway 2, because it nucleates at an angular position closer to the vertex, before the vortex has reached this angular position in the case of pathway 2. The difference arises from the different initial state, where for pathway 1 the chirality of the end vortex is the same as the overall chirality of the lower HR, while for pathway 2 it is opposite. By extracting the DW position from the frames of the movie for `HR4L\_pw2', as shown in \autoref{HR(S)1pw1pw2a_g}, the velocity of the field-driven DW could be determined from the initial slope to be $100\pm20$\,m/s. The maximum DW velocity in the corresponding simulation is during field application $230\pm60$\,m/s. The vortices in the simulations (HRS4) travel further than in the experiment. Pinning in the imperfect real samples explains the fact that the VW travels further in the simulations that do not incorporate structural defects that are inherent in the experiment. 

\section{Vortex wall chirality dependency on magnetic field direction and curvature}  

\begin{table}[H]
\begin{tabular}{p{0.1\textwidth}p{0.1\textwidth}|p{0.17\textwidth}p{0.17\textwidth}p{0.17\textwidth}p{0.17\textwidth}}
HR &   & $\vec{B}\uparrow \curvearrowright$       & $\vec{B}\downarrow \curvearrowright$       & $\vec{B}\uparrow \curvearrowleft$       & $\vec{B}\downarrow \curvearrowleft$        \\
\hline
1  & U &            &            & \textcolor{gray}{cw (H2H)}   & ccw (T2T)\text{*} \\
   & L &            &            & cw (H2H)\text{*} & \textcolor{gray}{ccw (T2T)}   \\ 
2  & U & \textcolor{gray}{\textit{ccw (H2H)}\text{*}} & cw (T2T)\text{*} &            &             \\
   & L &            &            & cw (H2H)\text{*} & \textcolor{gray}{\textit{ccw (T2T)}\text{*}}  \\
3  & U &            &            & \textcolor{gray}{cw (H2H)}   & ccw (T2T)\text{*} \\
   & L &            &            & cw (H2H)\text{*} & \textcolor{gray}{ccw (T2T)}   \\
4  & U & \textcolor{gray}{ccw (H2H)}  & cw (T2T)\text{*} &            &             \\
   & L &            &            & cw (H2H)\text{*} & \textcolor{gray}{ccw (T2T)}
 
\end{tabular}
\caption{Chiralities (cw: clockwise, ccw: counterclockwise) of the central VW after the first pulse (pw2, +y direction), after the second pulse (pw1, \textcolor{gray}{greyed out}); all measurements have been confirmed by micromagnetic simulations. $\vec{B}\uparrow$/$\vec{B}\downarrow$ indicates the field direction in relation to the curvature of the wire: $\curvearrowleft$/$\curvearrowright$ indicate the motion of the vortex in the HR element. U and L indicate upper and lower HR element. Results obtained by SEMPA measurements and micromagnetic simulations ($B=$\unit[15]{mT}) have normal font, Italic setting means only simulation with $B=$\unit[15]{mT}, an asterix indicates additional confirmation with B=24 mT simulation. “H2H” stand for head-to-head DW; “T2T” for tail-to-tail DW.}
\label{chirality}
\end{table}

In \autoref{chirality} the dependence of the chirality on field direction and curvature is shown. We show by the experimental SEMPA imaging and corresponding simulations that we can consistently obtain a particular DW chirality with a certain combination of curvature and magnetic field.

\section{Temperature-dependent switching pathways}
\label{Temperature-dependent switching pathways}

\begin{figure}[H]
\centering
	\includegraphics[width=0.69 \linewidth]{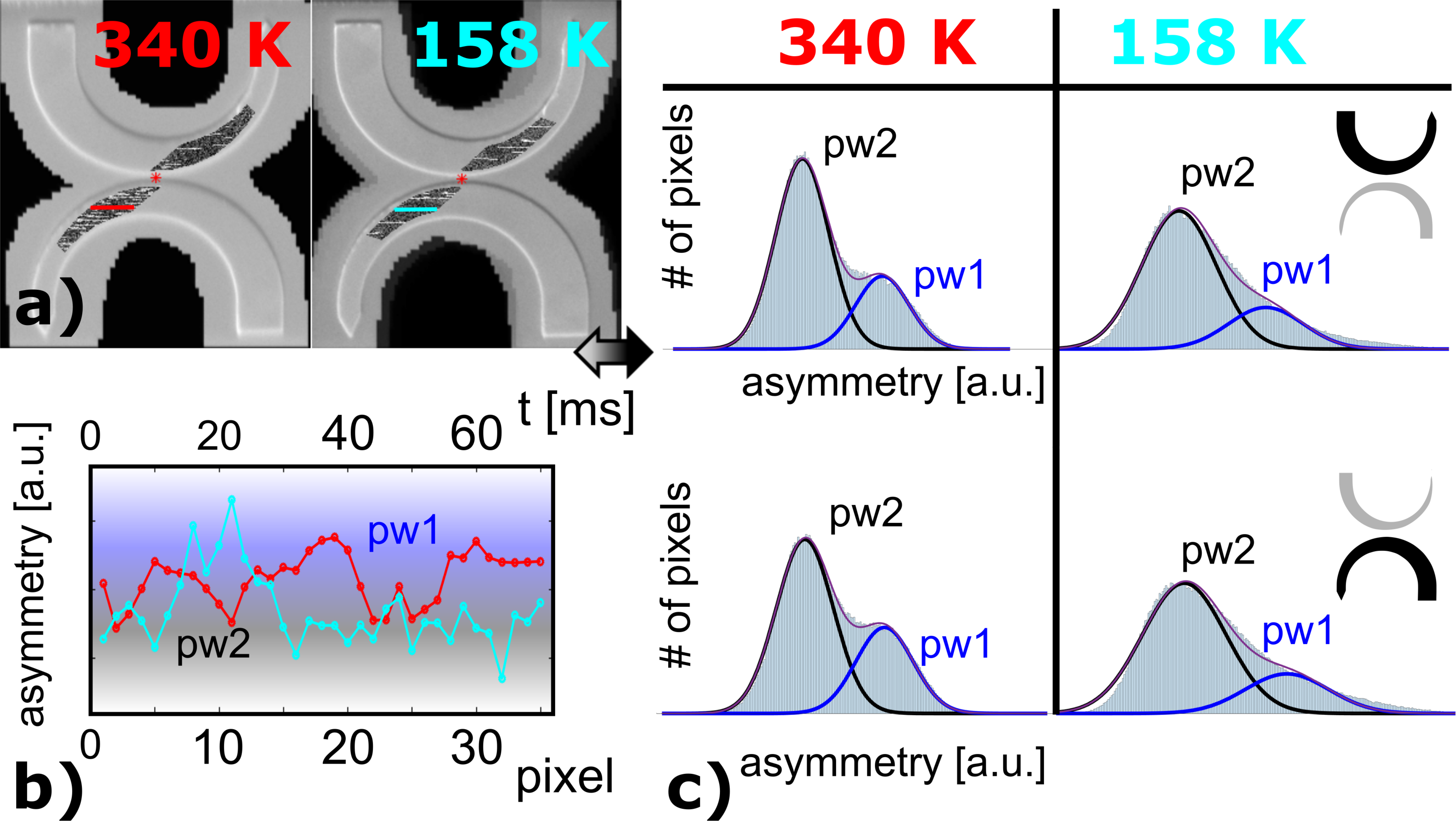}{\centering}
	\caption{Temperature dependence of the probability of the two pathways in HR3. \textbf{a)} Topographic image of HR3 with overlay of SEMPA single scan $x$ magnetization images of HR3 integrating over all adopted magnetic states during the whole excitation period at different temperatures. \textbf{b)} A line scan of the magnetization contrast as indicated by the lines in a). Note the different time scales as compared to the corresponding graph for HR4 in Fig. 3 in the main paper, since changing of the pathways occurs much more frequently in this arrangement. Blue background indicates asymmetry values corresponding to pathway 1 (pw1), black corresponds to pathway 2 (pw2). \textbf{c)} A double Gauss fit to the asymmetry values in the shown region of interest in a), revealing the overall relative occurrence of the two pathways.}
	\label{HR2_Tdependency}
\end{figure}

To identify the energy barrier relevant to the event responsible for changing the pathway we performed the measurements of HR3 at different temperatures. Due to Joule heating the sample heats up to \unit[340]{K} during the excitation when no sample cooling is employed (see methods section of the main paper). For HR3 the measurement was repeated for a cryostat temperature of \unit[158]{K}. The dynamics of the different pathways did not change significantly, but the relative frequency of the different pathways following the two pulses and therefore the frequency of changes between pathways was considerably different. In \autoref{HR2_Tdependency}a) single scans of the $x$-spin asymmetry are presented, overlayed on the topography image. The jumps in the contrast indicate the change of the magnetization direction associated with the change in the switching pathway, as shown in the line scans in \autoref{HR2_Tdependency}b). The dwell time at one pixel during the scan is \unit[2]{ms}, so that the full line scan covers a time span of \unit[70]{ms}. Integrated over 5.8\,billion excitation periods the pathway switching event occurred 2.2 times more often at \unit[340]{K} compared to \unit[158]{K}, indicating thermally excited depinning of the DW. From a histogram containing the pixel values of the SEMPA image in all acquired scans in the selected region of interest in the narrower part of the HR (\autoref{HR2_Tdependency}a)), two Gauss curves can be fitted to reveal the relative occurrence of both pathways, which can be estimated by comparing the area under the Gauss fit as shown in \autoref{HR2_Tdependency}c). It is revealed that pathway 1 occurs less often at \unit[158]{K} compared to \unit[340]{K}, or in other words pathway 2 is more stable.
The energy barrier which needs to be overcome to release the VW from the kinetic pinning potential can be calculated using the Arrhenius law $\tau=\tau_0\cdot\mathrm{exp}(\Delta E/k_B T)$. The attempt frequency $\tau_0$ which is about $10^{10}$\,Hz in magnetic systems~\cite{vouille2004thermally} can be assumed to be temperature-independent. The empirical probability $P$ for pathway switching was determined from the intensity jumps in the single scan asymmetry image ROIs (regions of interest) of HR3 at \unit[340]{K} and \unit[158]{K} for the upper HR `U' and the lower HR `L', which can be seen in \autoref{HR2_Tdependency}. The ratio of the rate constants $\tau$ for both temperatures can be set equal to the inverse ratio of the empirical probabilities $P$ for \unit[340]{K} and \unit[158]{K}. This in the end yields an energy barrier height of $3.1\times 10^{-21}$\,J for HR3 `U' and $3.0\times 10^{-21}$\,J for HR3 `L'. This corresponds to a temperature between 217\,K and 225\,K.

\section{Interaction of adjacent VWs}
By comparing the frequency of changes to the pathway, attributed to automotion, in HRs with axis symmetry and point symmetry, we see that the coupling of the two adjacent VWs plays a crucial role. In general if there are VWs in the HR center, there is always a head-to-head and tail-to-tail VW pair. Due to the opposite magnetic charge, the VWs have always a net attraction to each other~\cite{hayward2010pinning}, but there are smaller contributions to the energy depending on the stray field of the vortex cores and the half antivortices~\cite{tchernyshyov2005fractional} at the edges of the wires, which here play a role. The fact that the pathways of the upper and lower HR are coupled also can be explained by the stray field interaction of the DWs. The missing attractive interaction between the VWs in the period of excitation after one VW has experienced automotion increases drastically the probability for the single VW to undergo automotion, too.

\section{List of supplemental SEMPA and simulation videos}
In addition to the images shown in the main text we provide the full movies of both the SEMPA experimental data and the simulations. The SEMPA movies consist of three images showing the topography, the $x$ and the $y$ asymmetry. The simulations are colorcoded as in the main paper. One frame in the experimental SEMPA movies equals 2\,ns, in the simulation the frame rate is 1\,GHz. The experimental data is labeled HR1, HR2 etc., the simulations HRS1, HRS2, etc. The given time in the experimental SEMPA movies corresponds to the excitation graph provided in Fig.\,1c) in the main text. For the simulation videos the first field pulse starts with the first frame (as shown in the inset).
\begin{enumerate}
\item HR1 pw2 150\,ns (340\,K)
\item HR2 pw2 150\,ns (339\,K)
\item HR3 pw2 150\,ns (340\,K)
\item HR3 pw2 150\,ns (158\,K)
\item HR4 pw1 150\,ns (340\,K)
\item HR4 pw2 150\,ns (340\,K)
\item HRS1 pw2 109\,ns
\item HRS2 pw2 96\,ns
\item HRS3 pw2 113\,ns
\item HRS4 pw1 114\,ns
\item HRS4 pw2 90\,ns
\end{enumerate}

\bibliography{references}